\documentclass[twocolumn,pra,amsmath,amssymb,superscriptaddress]{revtex4-1}

\usepackage[pdftex]{graphicx}
\usepackage{dcolumn}
\usepackage{bm}
\usepackage{amsmath, amssymb,amsfonts}
\usepackage{mathrsfs}
\usepackage{physics}
\usepackage{type1cm}
\usepackage[colorlinks,linkcolor=blue,citecolor=blue]{hyperref}

\begin{document}

\title{
Two-dimensional correlation propagation dynamics with \\a cluster discrete phase-space method
}

\author{Kazuma Nagao}
\email{kazuma.nagao@riken.jp}
\affiliation{%
Computational Materials Science Research Team, RIKEN Center for Computational Science (R-CCS), Hyogo 650-0047, Japan
}%
\affiliation{%
Quantum Computational Science Research Team, RIKEN Center for Quantum Computing (RQC), Wako, Saitama 351-0198, Japan
}%
\author{Seiji Yunoki}
\affiliation{%
Computational Materials Science Research Team, RIKEN Center for Computational Science (R-CCS), Hyogo 650-0047, Japan
}%
\affiliation{%
Quantum Computational Science Research Team, RIKEN Center for Quantum Computing (RQC), Wako, Saitama 351-0198, Japan
}%
\affiliation{%
Computational Condensed Matter Physics Laboratory, RIKEN Cluster for Pioneering Research (CPR), Saitama 351-0198, Japan
}%
\affiliation{%
Computational Quantum Matter Research Team, RIKEN Center for Emergent Matter Science (CEMS), Saitama 351-0198, Japan
}%
\date{\today}

\begin{abstract}

Nonequilibrium dynamics of highly-controlled quantum systems is a challenging issue in statistical physics and quantum many-body 
physics, relevant to recent experimental developments of analog and digital quantum simulations.
In this work, we develop a discrete phase-space approach for general SU($N$) spin systems that utilizes cluster mean field equations, which capture non-trivial quantum correlations inside each cluster, beyond the capability of the standard discrete truncated Wigner approximation for individual classical spins. 
Our formalism, based on a cluster phase-point operator, enables efficient numerical samplings of cluster phase-space variables, where the total number of noise variables for a direct product state is independent of 
the specific way in which the entire system is divided into multiple equally sized finite clusters. 
We numerically demonstrate that the cluster discrete truncated Wigner approximation (C-dTWA) method can reproduce key results 
in a recent experiment on correlation propagation dynamics in a two dimensional Bose-Hubbard system.
We further compare the results of C-dTWA for clusters of $2\times 2$ sites with those from a two-dimensional tensor network method 
and discuss that both approaches agree very well in the short-time region, where the energy conservation is well maintained in the 
tensor network simulations. 
Since we formulate the 
C-dTWA method in a general form, it has the potential for application to various dynamical problems in isolated and open quantum 
systems, even in higher dimensions. 

\end{abstract}

\maketitle

\section{Introduction}
\label{sec: intro}

Recent experimental advances in analog and digital quantum simulations have stimulated exploration into 
nonequilibrium quantum dynamics of strongly interacting systems.
Typical experimental platforms that allow access to real-time quantum evolution in a highly controlled manner include 
ultracold atoms and molecules in optical lattices~\cite{trotzky2012probing,gross2017quantum,lepoutre2019out,schafer2020tools}, 
trapped ions~\cite{lanyon2011universal,blatt2012quantum}, 
optical tweezer arrays loaded with Rydberg atoms~\cite{kaufman2021quantum,bluvstein2021controlling,bluvstein2022quantum,bharti2023picosecond}, 
and superconducting qubits~\cite{kjaergaard2020superconducting,kim2023evidence}.
These platforms have been exploited to realize various types of quantum dynamics, 
such as 
quantum thermalization~\cite{trotzky2012probing,langen2016prethermalization,kaufman2016quantum}, 
nonergodic dynamics in many-body localization (MBL) regimes~\cite{choi2016exploring,bordia2017probing,luschen2017signatures}, 
correlation propagation dynamics following quenches~\cite{cheneau2012light,takasu2020energy}, 
discrete time crystalline states~\cite{zhang2017observation,frey2022realization,shinjo2024unveiling},
and superdiffusive magnetization dynamics in quantum circuits~\cite{rosenberg2023dynamics}.

Quantitative numerical methods are essential for advancing experimental techniques to control quantum dynamics and for testing the foundation of nonequilibrium statistical mechanics by cross-check against quantum simulation results.
However, simulating quantum dynamics on classical computers typically encounters significant challenges in many-body systems, 
such as the exponential growth of the Hilbert space dimension and the sign problem in continuous-time quantum Monte Carlo methods.
The tensor network method provides a versatile framework for reducing computational costs in simulating coherent quantum dynamics~\cite{schollwock2011density,cirac2021matrix,xiang2023density,mc2021stable,dziarmaga2023tensor}.
Despite its many advantages, the efficient application of the tensor network method is typically limited to low-entangled states 
in one-dimensional systems with short-range interactions. 
The phase space method offers an alternative approach for analyzing the quantum dynamics of many-body systems 
in a semiclassical manner~\cite{gardiner2004quantum,polkovnikov2010phase}. 
Among various representations, such as the positive-$P$ representation~\cite{steel1998dynamical,deuar2021fully,deuar2021multi} 
and the Wigner-Weyl-Moyal representation~\cite{blakie2008dynamics,polkovnikov2010phase,davidson2015s,nagao2019semiclassical,nagao2020fluctuations,huber2021phase,begg2024nonequilibrium}, the discrete truncated Wigner approximation (dTWA) method is a valuable scheme~\cite{schachenmayer2015many,orioli2018relaxation,lepoutre2019out,zhu2019generalized,signoles2021glassy,kunimi2021performance,mink2022hybrid,singh2022driven,muleady2023validating}, as it provides a natural semiclassical expression 
of finite-dimensional systems comprised of SU($N$) group operators.

The dTWA is formulated solely in terms of site-decoupled mean-field dynamics of spin variables, with initial conditions stochastically 
sampled from a positive probability distribution without a Gaussian approximation. 
Consequently, the computational cost remains polynomial as the system size or particle number increases. 
However, the accuracy is generally limited to short-time regimes due to the omission of higher-order quantum effects 
in this leading-order description. 
The performance evaluations for $N=2$ have been discussed for typical spin-1/2 models in Refs.~\cite{sundar2019analysis,kunimi2021performance}.
Additionally, in Ref.~\cite{nagao20213}, one of the authors of this paper has attempted to apply the SU(3) dTWA 
and the Gaussian SU(3) TWA to a quantum quench experiment on single-particle correlations in a two-dimensional strongly 
interacting Bose-Hubbard system~\cite{takasu2020energy}. 
However, it was noted that the results from these approaches showed noticeable deviations from the experimental results 
even at relatively short times.

Extension beyond the capability of the single-site dTWA has been explored in previous studies~\cite{pucci2016simulation,orioli2017nonequilibrium,kunimi2021performance}, based on an extended mean-field theory 
that exploits the Bogoliubov-Born-Green-Kirkwood-Yvon (BBGKY) hierarchy truncation.
However, this approach suffers from numerical instabilities in the time evolution~\cite{kunimi2021performance}. 
An alternative to the single-site description for spins is the cluster truncated Wigner approximation (CTWA)~\cite{wurtz2018cluster}.
This approach introduces cluster operators, each of which envelops multiple Pauli operators over a finite region in real space, 
in order to improve the accuracy of semiclassical approximations.
Nevertheless, the sampling of cluster variables suffers from several problems. 
For instance, a classical spin representation of the CTWA exploits a Gaussian approximation of the exact Wigner function. 
Such a method with an approximate Gaussian sampler is typically referred to as Gaussian CTWA.
The total number of Gaussian noise variables increases with the dimension of possible states in a cluster, leading to greater 
computational difficulty as the cluster size grows~\cite{wurtz2018cluster}.
Moreover, the construction of Gaussian Wigner functions is generally very complicated and hard to generalize to various situations, 
limiting the broader application of this promising method.
Therefore, it is highly desirable to develop an approach that can diminish these difficulties in the numerical sampling 
by utilizing techniques in the dTWA.

In this paper, we present a general framework that combines the CTWA approach and the sampling technique of the SU($N$) dTWA.
We refer to this approach as the cluster discrete TWA (C-dTWA). 
Our formulation is based on {\it dynamical cluster phase-point operators}, defined for an arbitrary set of clusters.
One advantage of this approach is that it enables efficient numerical sampling for fluctuating cluster variables 
with positive definite probabilities.
This efficiency arises from the fact that the total number of initial fluctuations depends only on the total system size and is 
independent of the cluster size.  
Consequently, the numerical sampling is easily scalable, regardless of the specific clustering choice. 
We specifically apply the C-dTWA to the correlation propagation dynamics in a two-dimensional Bose-Hubbard system~\cite{takasu2020energy,nagao20213,kaneko2022tensor} and demonstrate that this method significantly improves 
the accuracy of the previous SU(3) dTWA. 
In addition, we assess the performance of the C-dTWA in comparison with the Gaussian CTWA, experimental results from Ref.~\cite{takasu2020energy}, and corresponding tensor-network results based on the infinite projected entangled pair states (iPEPS) method~\cite{kaneko2022tensor}.

The rest of this paper is organized as follows.
In Sec.~\ref{sec: formulation}, we formulate the C-dTWA for a generic system comprised of SU($N$) spins.
In Sec.~\ref{sec: correlation}, we apply this method to a strongly interacting Bose-Hubbard system and demonstrate 
the performance of C-dTWA in comparison with experiments and other numerical methods.
In Sec.~\ref{sec: conclusions}, we conclude the paper with future perspectives. 
In Appendix~\ref{app: ising}, we provide additional results comparing the C-dTWA and Gaussian CTWA methods. 
In Appendix~\ref{app: manual}, we supplement details of the C-dTWA and Gaussian CTWA formulations for a state-restricted 
Bose-Hubbard model.

\section{Formulation: Cluster-dTWA for SU($N$) spin systems}
\label{sec: formulation}

For concreteness, we specifically consider a generic SU($N$) spin system isolated from the environment, 
described by the following Hamiltonian:  
\begin{align}
{\hat {\cal H}} = \sum_{i,j,a,b} J_{i,j}^{a,b} {\hat s}^{a}_{i}{\hat s}^{b}_{j} + \sum_{i,a}h_{i}^{a}{\hat s}^{a}_{i}, \label{eq: hamiltonian_general}
\end{align}
where $i$ and $j$ are site indices in real space, and $a$ and $b$ are internal indices on spin space. 
The dynamics of the quantum states are determined by the commutation relations for the spin operators 
given by  $[ {\hat s}^{a}_{j}, {\hat s}^{b}_{j'}] = i \sum_{c=1}^{N} \epsilon_{abc} {\hat s}^{c}_{j}\delta_{j,j'}$. 
The spin operators ${\hat s}^{a}_{j}$ are Hermitian and traceless, satisfying 
$({\hat s}^{a}_{j})^{\dagger} = {\hat s}^{a}_{j}$ and ${\rm Tr}{\hat s}^{a}_{j} = 0$.
For convenience in later use, we introduce the set $V = \{1,2,\cdots, N_{\rm s} \}$, where $N_{\rm s}$ is the total number of sites.

\begin{figure}
\begin{center}
\includegraphics[width=75mm]{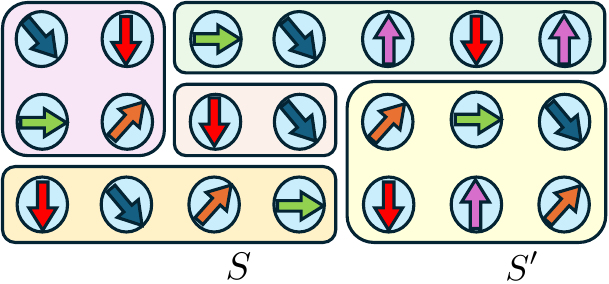}
\caption{
Clustering an SU($N$) spin system with a set of finite regions.
Regions such as $S$ and $S'$ are introduced to define cluster operators that encompass individual spins. 
In the C-dTWA, each region has a time-dependent cluster phase-point operator that evolves according to 
a cluster mean-field equation. 
}
\label{fig: clusters}
\end{center}
\end{figure}

Let us define clusters of individual SU($N$) spins, as illustrated in Fig.~\ref{fig: clusters}.
A key component of our formulation is the cluster phase-point operator $ {\hat {\cal A}}_{S_\mu} = {\hat A}_{j_1} \otimes {\hat A}_{j_2} \cdots  \otimes {\hat A}_{j_l} \in {\cal H}^{\otimes l }$ on a finite region $S_\mu$, where $\mu$ labels the cluster, 
$\{ j_1,j_2,\cdots,j_l \} = S_{\mu} $, ${\cal H}$ is the local Hilbert space, and $l = {\rm dim} (S_\mu)$ is the cluster size. 
This operator provides a phase-space representation of the time-dependent density matrix operator 
${\hat \rho}(t) = {\hat U} (t,0) {\hat \rho}(t=0) {\hat U}^{\dagger}(t,0)$, expressed as 
\begin{align}
{\hat \rho}(t) = {\hat U}(t,0) \overline{(\bigotimes_{\mu \in V_{\rm c}} {\hat {\cal A}}_{S_\mu} )}{\hat U}^{\dagger}(t,0),
\label{eq:rho_t}
\end{align}
where $V_{\rm c}$ is the set of cluster indices. 
In the single-site dTWA, $V_{\rm c} = V$ holds.
In this work, we assume that ${\hat \rho}(t = 0)$ is a factorized operator over clusters, e.g., a direct product pure state. 
The overline in Eq.~(\ref{eq:rho_t}) 
denotes an ensemble average over different configurations of onsite classical spins $r^{a}_{i} \in {\mathbb R}$, 
while the direct sum of $S_{\mu}$ covers $V$, i.e., $\cup_{\mu \in V_{\rm c} } S_{\mu} = V$.
The onsite operator ${\hat A}_{j}$ is expanded as ${\hat A}_{j} = N^{-1} {\hat {\mathbb I}}_{j} + g^{-1}\sum_{a} r^{a}_{i} {\hat s}^{a}_{i}$, 
where $g = {\rm Tr}({\hat s}^{a}_{i})^2$ is a normalization factor.
The fluctuations of $r^{a}_{i}$ are sampled according to a set of positive probability distributions 
derived from a quantum-state-tomography method for SU($N$) systems,  
generalizing the state-of-the-art sampling method in the generalized dTWA (GDTWA)~\cite{zhu2019generalized}.
Hence, ${\hat \rho}(t=0)$ can be reconstructed as the ensemble average of {\it measurement outcomes of local spin configurations}.

Next, we approximate the time evolution of ${\hat \rho}(t)$ in a numerically tractable manner.
Within the dTWA in the cluster representation, the density matrix operator at time $t$ is approximated as 
${\hat \rho}(t) \approx \overline{\bigotimes_{\mu \in V_{\rm c}} {\hat {\mathscr A}}^{\rm cl}_{S_\mu}(t)}$.
The cluster phase-point operators ${\hat {\mathscr A}}_{S_\mu}^{\rm cl}(t)$ are updated step by step 
according to a cluster mean-field equation: 
\begin{align}
\hbar\partial_t {\hat {\mathscr A}}^{\rm cl}_{S_\mu}(t) = - i [{\hat \Gamma}^{\rm intra}_{S_\mu} + {\hat \Phi}^{\rm inter}_{S_\mu}, {\hat {\mathscr A}}^{\rm cl}_{S_\mu}(t)]. \label{eq: cluster_mf}
\end{align}
For the Hamiltonian operator in Eq.~(\ref{eq: hamiltonian_general}), intra-cluster correlations inside each cluster are described 
by ${\hat \Gamma}^{\rm intra}_{S_\mu} = \sum_{i \in S_{\mu}}\sum_{a}h_{i}^{a}{\hat s}^{a}_{i} + \sum_{i,j \in S_{\mu}}\sum_{a,b} J_{i,j}^{a,b} {\hat s}^{a}_{i}{\hat s}^{b}_{j}$.
In contrast, inter-cluster couplings reduce to ${\hat \Phi}^{\rm inter}_{S_\mu} = \sum_{i \in S_{\mu}} \sum_{\nu\, (\neq \mu)} \sum_{j \in S_{\nu}} \sum_{a,b} J_{i,j}^{a,b} {\hat s}^{a}_{i} \langle {\hat s}^{b}_{j} \rangle_{S_{\nu}}$.
Here, $\langle {\hat s}^{b}_{j} \rangle_{S_{\nu}} =  {\rm Tr} \left[ {\hat s}^{b}_{j} {\hat {\mathscr A}}^{\rm cl}_{S_\nu}(t) \right]$ are 
regarded as the {\it cluster mean fields} coupled to the cluster on $S_{\mu}$.

In numerical simulations, the state of ${\hat {\mathscr A}}^{\rm cl}_{S_\mu}(t)$ is represented as a set of finite dimensional matrices.
With a randomly drawn set of $r^{a}_{i}$ across the entire system, $\bigotimes_{\mu \in V_{\rm c}} {\mathscr A}^{\rm cl}_{S_\mu}(t) $ 
forms an $N^{N_{\rm s}}$ dimensional random matrix, whose ensemble average no longer retains a direct product structure 
for $t > 0$.
This feature reflects the dynamic growth of inter-cluster correlations~\cite{wurtz2018cluster,kunimi2021performance}.
The intra-cluster correlations are fully incorporated along the trajectory dynamics beyond the capability of the single-site dTWA. 
The benefit of this sampling method is that the total number of random variables, i.e., $r^{a}_{i}$, remains independent of 
the cluster choice. 
This can be contrasted with the Gaussian CTWA, where  
a Gaussian Wigner function, even simplified with a dimensional reduction technique for an appropriate operator basis of clusters,  
involves $2 D$ fluctuating real variables, with $D$ being the cluster size~\cite{wurtz2018cluster}.
Consequently, the total number of variables increases with $D$, which becomes substantial for large clusters ($D \gg 1$).
In Appendix~\ref{app: ising}, we numerically demonstrate that the C-dTWA can reproduce results consistent with those obtained 
by the Gaussian CTWA for an SU(2) transverse field Ising model, indicating both the consistency and advantage of 
the C-dTWA method. 
More interestingly, for nonequilibrium dynamics of the two-dimensional Bose-Hubbard model, which is the primary focus of this work, 
the C-dTWA demonstrates higher accuracy than the Gaussian CTWA in evaluating the observables studied here.
Further details are discussed in Sec.~\ref{sec: correlation}.

\section{Correlation propagation dynamics in two dimensions}
\label{sec: correlation}

To demonstrate the effectiveness of using correlated clusters in the dTWA, we specifically analyze the correlation propagation dynamics 
in an isolated Bose-Hubbard model on a square lattice~\cite{takasu2020energy,kaneko2022tensor}. 
The Hamiltonian is given by 
\begin{align}\label{eq:H_bose}
{\hat H} = -J\sum_{\langle i,j \rangle}({\hat a}^{\dagger}_{i}{\hat a}_{j}+{\rm H.c.}) + \frac{U}{2}\sum_{i}{\hat a}^{\dagger}_{i}{\hat a}^{\dagger}_{i}{\hat a}_{i}{\hat a}_{i},
\end{align}
where $\hat a_i^\dag$ ($\hat a_i$) is the creation (annihilation) operator of bosons at site $i$, 
and $\langle i,j \rangle$ denotes pairs of nearest-neighbor sites. 
The hopping amplitude $J$ and the onsite interaction $U$ are tunable via the isotropic lattice depth $\mathcal{V}$.
In quench experiments, the dynamics is observed in a strongly interacting regime where $U \gg J$.
Under this condition, the relevant Hilbert space is restricted, and here we set  
the maximum occupation to $n_{\rm max} = 2$.
Consequently, any local operator can be represented using a three-dimensional matrix, which can be decomposed in base matrices 
of the SU(3) group~\cite{davidson2015s,nagao20213}.
This state truncation transforms the original Hamiltonian into a model of interacting SU(3) pseudospins, representing a specific 
case of the general model in Eq.~(\ref{eq: hamiltonian_general}). 
Technical details of the C-dTWA for this reduced system, as well as those of the Gaussian CTWA, are provided in 
Appendix~\ref{app: manual}.

\begin{figure}
\begin{center}
\includegraphics[width=85mm]{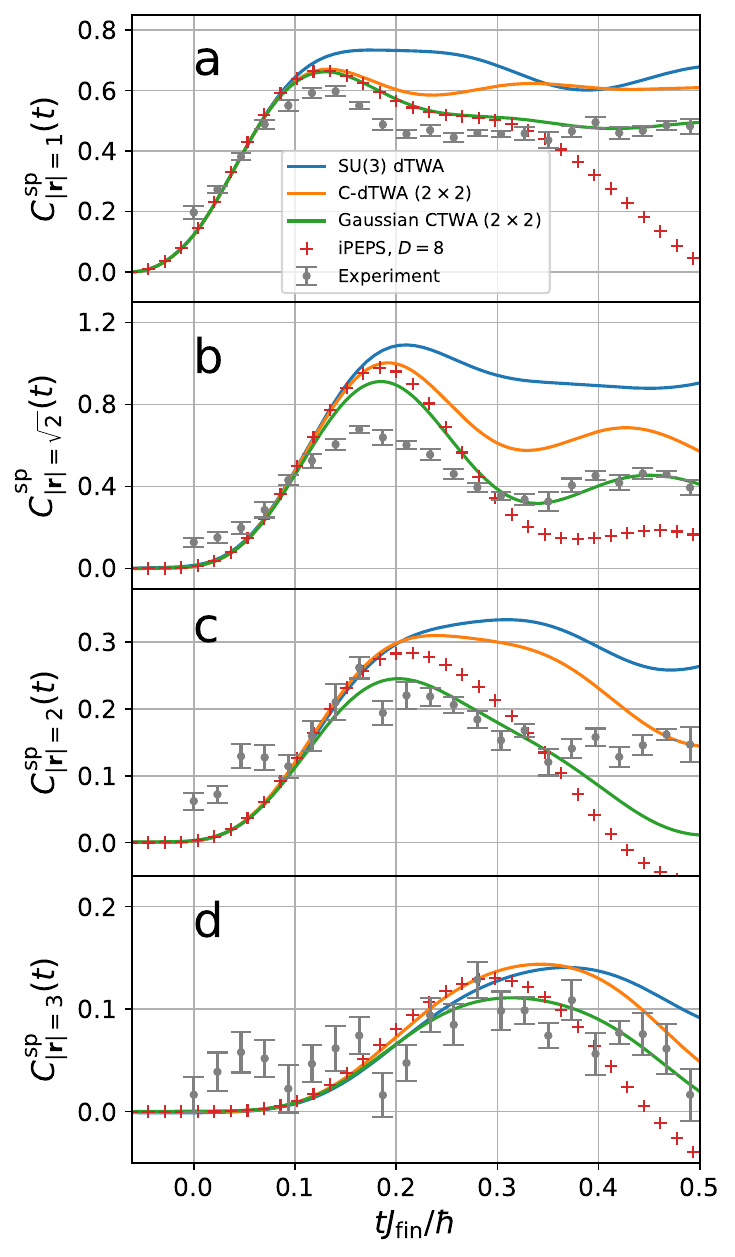}
\vspace{-6mm}
\caption{
Time evolution of the single-particle correlation functions $C^{\rm sp}_{|{\bf r}|}(t)$ after a finite-speed ramp down of the optical lattice 
depth (completed at $t=0$) at distances (a) $|{\bf r}|=1$, (b) $|{\bf r}| = \sqrt{2}$, (c) $|{\bf r}| = 2$, and (d) $|{\bf r}| = 3$, 
evaluated using the C-dTWA with clusters of $2\times 2$ sites (orange line), the Gaussian CTWA with the same clusters (green line), and  the SU(3) dTWA (blue line). 
The total number of sites is $N_{\rm s} = 400$ with $L_x = L_y = 20$ under periodic boundary conditions. 
For comparison, the results for the iPEPS with the bond dimension $D=8$ (red crosses)~\cite{kaneko2022tensor} 
and the experiments (gray points with error bars)~\cite{takasu2020energy} are also plotted. 
}
\label{fig: exp}
\end{center}
\end{figure}

\begin{figure}
\begin{center}
\includegraphics[width=80mm]{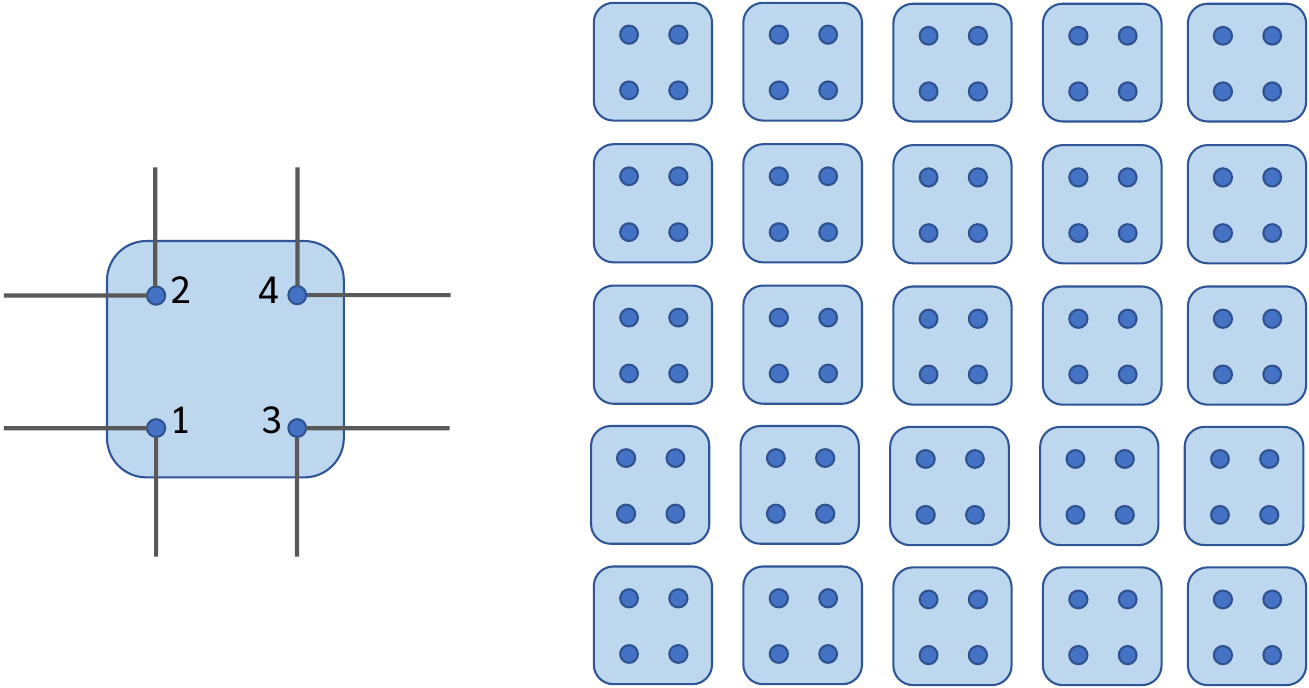}
\caption{
Cluster phase-space representation of the two-dimensional Bose-Hubbard system on a square lattice with clusters of $2\times 2$ sites.
The indexing rule within a cluster is shown on the left, where the outer eight legs represent nonlinear couplings to neighboring clusters. 
}
\label{fig: bhm2by2}
\end{center}
\end{figure}

We analyze a quantum quench starting from a direct product state.
The relevant quantity is the equal-time single-particle correlation function, defined as 
$C^{\rm sp}_{(\varDelta_x,\varDelta_y)}(t) = N_{\rm s}^{-1}\sum_{i,j}' \langle {\hat a}^{\dagger}_{i}(t) {\hat a}_{j}(t) \rangle$, 
where $\sum'_{i,j}$ indicates the summation 
over sites $i$ and $j$ with the conditions $|x_i-x_j|=\varDelta_x$ and $|y_i-y_j|=\varDelta_y$~\cite{nagao20213}, 
and $(x_i,y_i)$ represents the coordinates of site $i$. 
The propagation of correlations is measured in the units of the Euclidean distance $|{\bf r}| = \sqrt{\varDelta_x^2 + \varDelta_y^2}$.
Prior to the quench at $t=-\tau_{\rm Q}$, the system is prepared in the ground state for $U = \infty$, 
which corresponds to the unit-filling Mott insulating state $\ket{\psi_{\rm ini}} = \bigotimes_{i=1}^{N_{\rm s}}\ket{1}_{i}$.
Subsequently, the optical lattice depth $s(t) = \mathcal{V}/E_{\rm R}$ is linearly varied over a finite ramp time $\tau_{\rm Q}$, 
where $E_{\rm R}$ is the recoil energy of the system. 
The time evolution involves ramping down the lattice, i.e., $s (t) = \frac{t+\tau_{\rm Q}}{\tau_{\rm Q}}(s_{\rm f}-s_{\rm i}) + s_{\rm i}$ 
for $t$ from $-\tau_{\rm Q}$ to $0$, 
with the initial and final lattice depths of $s_{\rm i}=15$ and $s_{\rm f}=9$, respectively.
This final lattice depth corresponds to $U/J = 19.6$.
The ramp time is set to $\tau_{\rm Q} = 0.1$~ms, which is sufficiently rapid~\cite{nagao20213,kaneko2022tensor}.
In the following analyses, we assume periodic boundary conditions. The lattice geometry consists of a total of 
$N_{\rm s}=L_x\times L_y=400$ sites with $L_x = L_y = 20$.

Figure~\ref{fig: exp} shows the equal-time single-particle correlation functions for $|{\bf r}|=1$, $\sqrt{2}$, $2$, and $3$, 
calculated using the C-dTWA with $2 \times 2$ site clusters. These results are compared with those obtained from 
the iPEPS~\cite{kaneko2022tensor}, the SU(3) dTWA, and the experiment~\cite{takasu2020energy}.
The sampling scheme for the SU(3) dTWA is based on the quantum-state-tomography method 
in the GDTWA~\cite{zhu2019generalized,nagao20213}.
The cluster configuration used in the C-dTWA is illustrated in Fig.~\ref{fig: bhm2by2}.
Figure~\ref{fig: exp}(a) shows the results for the nearest-neighbor correlation function 
$C^{\rm sp}_{|{\bf r}|=1} (t)= (C^{\rm sp}_{(1,0)} (t) + C^{\rm sp}_{(0,1)}(t) )/2$. 
The first peak appears around $t \sim 0.12 \hbar/J_{\rm fin}$ in the iPEPS, C-dTWA, and experimental results. 
The C-dTWA and iPEPS results are in good agreement for $t \lesssim 0.2 \hbar/J_{\rm fin}$,  
where $J_{\rm fin}$ denotes the final hopping amplitude after the quench. 
The peak time obtained by these numerical methods is consistent with the experimental observation. 
However, the SU(3) dTWA fails to capture this peak because of the breakdown of the method for $t \gtrsim 0.1 \hbar/J_{\rm fin}$, 
as discussed in the previous work~\cite{nagao20213}.
In the iPEPS calculations, the energy is approximately conserved for $t \lesssim 0.4 \hbar/J_{\rm fin}$~\cite{kaneko2022tensor}.
Even at longer times, for $t \gtrsim 0.4 \hbar/J_{\rm fin}$, where the iPEPS is no longer validated, the C-dTWA results remain 
qualitatively consistent with the experimental behavior.

Figure~\ref{fig: exp}(b) shows the results for $C^{\rm sp}_{|{\bf r}|=\sqrt{2}} (t)= C^{\rm sp}_{(1,1)} (t)$. 
The correlation functions calculated using the C-dTWA and iPEPS also show good agreement for $t \lesssim 0.2 \hbar/J_{\rm fin}$, 
and the first peak time is consistent with the experimental observation.
However, the corresponding peak in the SU(3) dTWA appears slightly shifted to a later time. 
Additionally, the C-dTWA captures a second peak at $t \sim 0.43 \hbar/J_{\rm fin}$, which is qualitatively 
consistent with the experimental results.  
Figures~\ref{fig: exp}(c) and \ref{fig: exp}(d) display the correlation functions for $|{\bf r}|=2$ and $3$, 
respectively. These results 
demonstrate that the intra-cluster fluctuations treated by the C-dTWA, which may still be somewhat underestimated, 
improve the first peak time compared to the SU(3) dTWA, when the results are compared with those of the iPEPS and the experiment.
However, there remain slight deviations in the peak times between the C-dTWA and iPEPS results. 
To fully understand the origin of these deviations, it would be necessary to use larger clusters in the calculations.  
Such an analysis is beyond the scope of this study due to computational limitation 
and is left for future study.

Furthermore, as shown in Fig.~\ref{fig: exp}, the Gaussian CTWA results, obtained using the same $2\times 2$ site clusters 
as in the C-dTWA, appear to capture the overall features at and around the first peaks reasonably well, 
performing better in this regard than the single-site SU(3) dTWA. 
In particular, the results for the nearest-neighbor correlation $C^{\rm sp}_{|{\bf r}|=1}(t)$ show closer agreement with the iPEPS results 
up to $t \sim 0.3 \hbar/J_{\rm fin}$ and aligns more closely with the experimental results in the time region  
$0.4 \lesssim t J_{\rm fin}/\hbar \lesssim 0.5$.
However, as discussed below, this seems to be coincidental in the Gaussian approach. 
Figures~\ref{fig: exp}(b)--(d) indicate that the Gaussian CTWA becomes less accurate for distant correlations at $|{\bf r}| > 1$, 
as it fails to reproduce the height of the first peak, which is noticeably suppressed compared to the values calculated by 
the iPEPS and C-dTWA.
The long-time behavior after the peak time is also rather different from the C-dTWA results due to inherent limitations in  
the Gaussian sampling scheme. 
These two methods differ not only in the time evolution of the expectation values, estimated through the ensemble 
average of trajectories, but also in the sampling efficiency.
The Gaussian CTWA requires $160$ independent noises per cluster, introduced after diagonalizing intra-cluster correlations, 
whereas the C-dTWA requires only $16$. 
The scaling of the latter (former) scheme is polynomial (exponential) in cluster size.
Therefore, the C-dTWA offers advantages in reducing the computational cost of Monte Carlo sampling.

\begin{figure}
\begin{center}
\includegraphics[width=85mm]{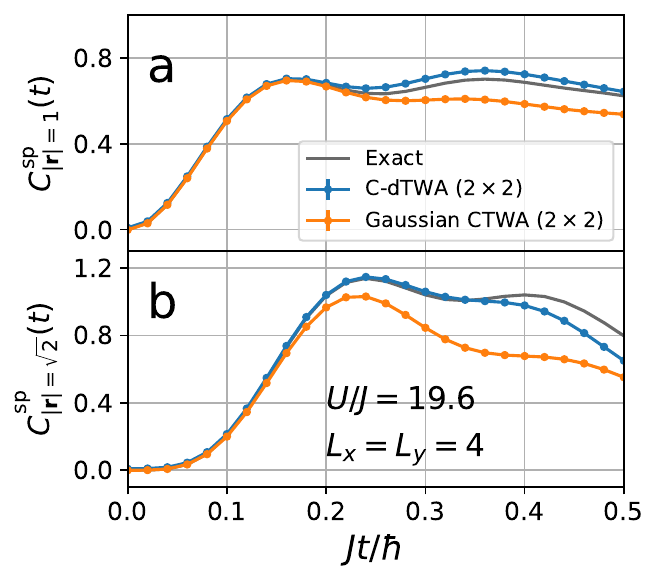}
\vspace{-6mm}
\caption{
Time evolution of the single-particle correlation functions $C^{\rm sp}_{|{\bf r}|}(t)$ 
at distances (a) $|{\bf r}|=1$ and (b) $|{\bf r}|=\sqrt{2}$ 
after a sudden quench at $t=0$ for $N_{\rm s}=16$ with $L_x = L_y = 4$ under periodic boundary conditions.
The initial state is set to be the unit-filling Mott insulating state and the final state after the quench corresponds to $U/J = 19.6$. 
The blue and orange dots represent the results for the C-dTWA and Gaussian CTWA with clusters of $2 \times 2$ sites, respectively. 
For comparison, the numerically exact results obtained using QuSpin library~\cite{weinberg2017quspin,weinberg2019quspin} are 
also shown as black solid curves.
}
\label{fig: exact}
\end{center}
\end{figure}

To gain insights on the quantitative limitations of the C-dTWA and Gaussian CTWA methods, we compare 
the correlation propagation dynamics obtained by these methods
with the numerically exact results for a small system of  $N_{\rm s}= 16$ sites with $L_x = L_y = 4$ under periodic boundary conditions.
As in the previous case, the initial state is set to be the unit-filling Mott insulating state.
For simplicity, here we assume $\tau_{\rm Q} = 0$, implying a sudden quench from $U/J = \infty$ to $U/J = 19.6$ at $t=0$.
Figure~\ref{fig: exact} shows the single-particle correlation functions $C^{\rm sp}_{|{\bf r}|}(t)$ at 
$|{\bf r}|=1$ and $\sqrt{2}$ calculated using the C-dTWA with clusters of $2 \times 2$ sites. 
The C-dTWA method successfully captures the first peaks at early times. 
However, for $|{\bf r}|=1$ ($|{\bf r}|=\sqrt{2}$), noticeable deviations from the exact results appear when  
$t \gtrsim 0.25 \hbar/J$ ($t \gtrsim 0.35 \hbar/J$), due to the higher order contributions that become relevant 
in the long time quantum dynamics.
The Gaussian CTWA method with the same clusters of $2 \times 2$ sites also correctly captures the first peaks for both cases. 
However, its accuracy for $|{\bf r}|=1$ $(|{\bf r}|=\sqrt{2}$) deteriorates  
as $t \gtrsim 0.25 \hbar/J$ ($t \gtrsim 0.15 \hbar/J$),  
and the first peak height for $|{\bf r}|=\sqrt{2}$ is remarkably lower than the exact value. 
The superior performance of the C-dTWA over the Gaussian CTWA can be attributed to the discrete nature of sampling 
in the C-dTWA, highlighting the strength of this approach.

These comparisons demonstrate that intra-cluster correlations are crucial for describing key features 
in the correlation propagation dynamics, such as the first peak times and the qualitative behavior following those peaks.
Moreover, the sampling scheme in the C-dTWA offers a distinct advantage, yielding better predictions specifically 
for $|{\bf r}| > 1$, compared with the Gaussian scheme used in the standard CTWA approach. 
Nevertheless, as shown in Fig.~\ref{fig: exact}, where the results are compared with the numerically exact calculations, 
the C-dTWA method with 
clusters of up to $2 \times 2$ sites does not fully capture the quantum dynamics, 
particularly at later stages. 
In principle, the C-dTWA method becomes asymptotically exact as the cluster size approaches the total system size. 
However, although the initial state sampling is simplified with a polynomial number of noises in the C-dTWA, 
the number of dynamical variables per trajectory still increases exponentially with the cluster size. 
This exponential scaling poses a practical limit on the cluster size, constrained by available computational resources, 
making it challenging to progress to larger clusters, such as $3\times3$ sites. 
In future work, we aim to develop a nontrivial parallelization approach combined with data compression techniques 
to increase the maximum feasible cluster size by exploiting modern large-scale supercomputer systems.

\section{Conclusions}
\label{sec: conclusions}

We introduced the C-dTWA method, which combines the cluster phase space representation method 
with the discrete truncated Wigner sampling method, 
for finite-dimensional SU($N$) spin systems.
To demonstrate the effectiveness of the method, we applied the C-dTWA to study the correlation propagation dynamics 
in a two-dimensional Bose-Hubbard model.
The C-dTWA results with clusters of $2\times 2$ sites agree with those of the iPEPS calculations at short times, 
where the energy conservation in the iPEPS calculations is maintained. 
Even at longer times, the C-dTWA successfully captures the qualitative behavior in the experiments, which is not captured 
in the iPEPS calculations.

While this study primarily focuses on assessing the performance of the C-dTWA method, it also paves 
the way to exploring nontrivial dynamics beyond the current capability of other numerical methods, such as tensor network methods 
and the singe-site dTWA. 
For instance, our method can evaluate long-distance correlations that are difficult to treat in the iPEPS approach,  
which is particularly valuable for determining propagation velocities from the dynamics of correlation functions, offering a more 
accurate estimation of the Lieb-Robinson bound for quantum information propagation.
Moreover, the cluster phase-point operator formalism developed here for individual SU($N$) spins is highly versatile,  
as it can be directly extended to open quantum systems described by the Lindblad master 
equation~\cite{nagao2023semiclassical}. These potential applications will be investigated in future work.

{\it Note added---}
While completing this paper, we became aware of a recent related study on a cluster generalization of 
the generalized discrete truncated Wigner approximation (GDTWA)~\cite{alaoui2024measuring}.
Although our approach shares a similar conceptual basis, there are notable differences. 
First, our formulation is based on a cluster phase-point operator, which enhances practicality in computations, 
whereas their approach is a straightforward generalization of the GDTWA for SU($N$) systems.
Second, their main application focuses on an experiment with spin-$3$ chromium atoms in a three-dimensional optical lattice, 
treating a neighboring pair of $s=3$ spins as an SU($49$) spin system.
In contrast, our study applies the method to a two-dimensional Bose-Hubbard system, where we treat a subsystem with $2\times 2$ 
sites as an SU($81$) spin system.

\begin{acknowledgments}

We thank Ryui Kaneko, Rongyang Sun, and Ippei Danshita for their valuable discussions, and 
Yosuke Takasu and Yoshiro Takahashi for providing experimental data in Ref.~\cite{takasu2020energy}.
This work is partially supported by the project JPNP20017, funded by the New Energy and Industrial Technology 
Development Organization (NEDO), Japan. We acknowledge the support from JSPS KAKENHI (Grant No. JP21H04446) 
from the Ministry of Education, Culture, Sports, Science and Technology (MEXT), Japan, as well as  
the funding received from 
JST COI-NEXT (Grant No. JPMJPF2221) and  
the Program for Promoting Research of the Supercomputer Fugaku (Grant No. MXP1020230411) from MEXT, Japan.  
Additional support was provided by 
the RIKEN TRIP initiative (RIKEN Quantum) and
the COE research grant in computational science from Hyogo Prefecture and Kobe City 
through the Foundation for Computational Science. 
The truncated Wigner simulations have been carried out on the HOKUSAI supercomputing system at RIKEN and the supercomputer 
Fugaku provided by the RIKEN Center for Computational Science. 
The numerically exact calculations have been done using iTensor~\cite{itensor} and 
QuSpin~\cite{weinberg2017quspin,weinberg2019quspin} packages. 

\end{acknowledgments}

\appendix

\section{Revisit to a four site transverse-field Ising chain}
\label{app: ising}

In this appendix, we revisit the benchmark of the cluster phase space description, which was also discussed in 
Ref.~\cite{wurtz2018cluster}, for a system of four coupled Pauli spins described by the following Hamiltonian: 
\begin{align}
{\hat H}_{\rm Ising} = J\sum_{i=1}^{3}{\hat \sigma}^{(i)}_{z}{\hat \sigma}^{(i+1)}_{z} + h_x \sum_{i=1}^{4}{\hat \sigma}^{(i)}_{x}. 
\label{eq:Hising}
\end{align}
Here, $\hat\sigma_\gamma^{(i)}$ is the $\gamma$ component of the Pauli matrices at site $i$ with $\gamma=x,y$, and $z$ . 
To implement the C-dTWA, we divide the system into two subsystems and 
define the global phase-point operator as ${\hat {\cal A}}_{\rm tot} = {\hat {\cal A}}_{L} \otimes {\hat {\cal A}}_{R}$.
The left and right cluster phase-point operators, ${\hat {\cal A}}_{L}$ and ${\hat {\cal A}}_{R}$, are respectively decomposed as follows: 
\begin{align}
{\hat {\cal A}}_{L} 
&= \frac{1}{2^2}\sum_{m,n \in \{0,x,y,z\}} X_{m,n}^{(1,2)}{\hat \sigma}_{m}^{(1)} \otimes {\hat \sigma}_{n}^{(2)}
\end{align}
and 
\begin{align}
{\hat {\cal A}}_{R} 
&= \frac{1}{2^2}\sum_{m,n \in \{0,x,y,z\}} X_{m,n}^{(3,4)}{\hat \sigma}_{m}^{(3)} \otimes {\hat \sigma}_{n}^{(4)},
\end{align}
where $\hat\sigma_0^{(i)}$ denotes the identity matrix, and $X_{0,0}^{(1,2)} = X_{0,0}^{(3,4)} = 1$ because the dynamics is unitary.

\begin{figure}
\begin{center}
\includegraphics[width=85mm]{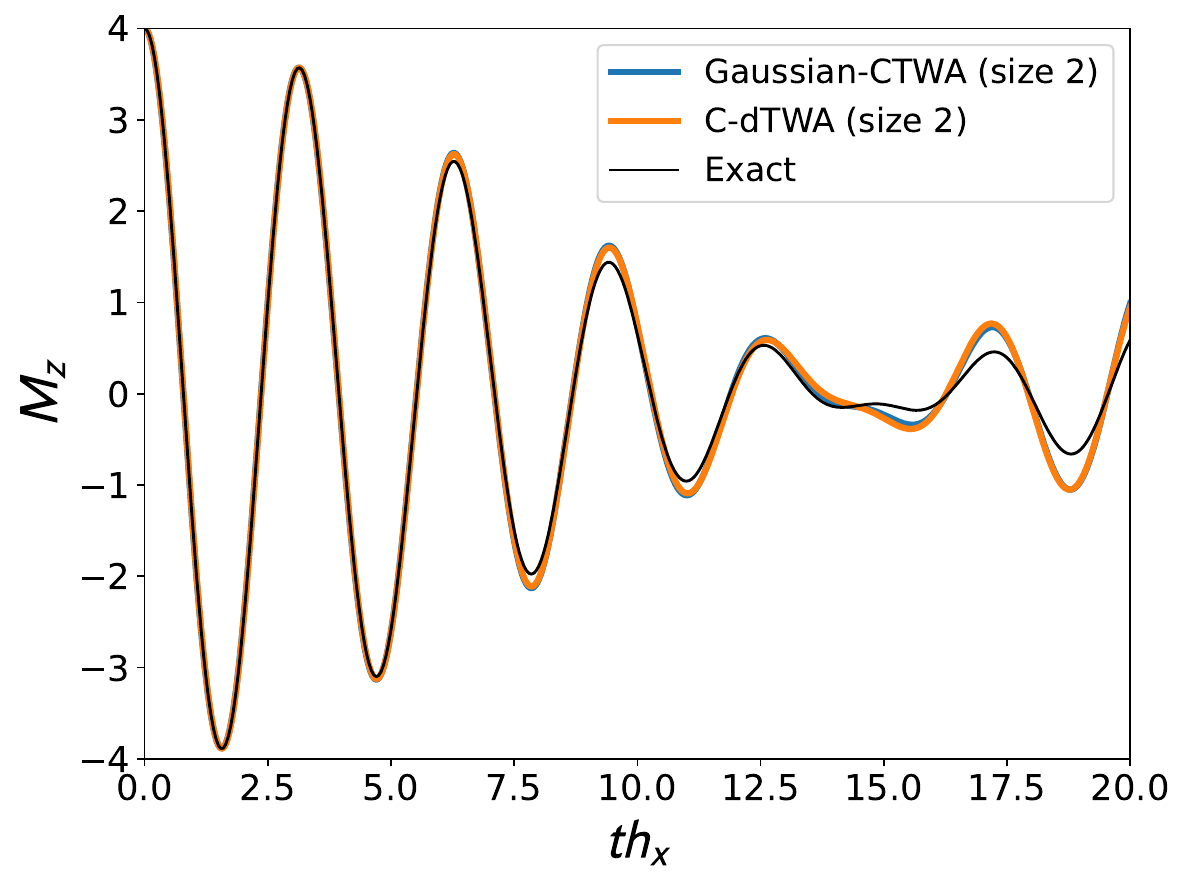}
\vspace{-3mm}
\caption{
Time evolution of the magnetization $M_z$ for the four site Ising chain, as described by the Hamiltonian in Eq.~(\ref{eq:Hising}), 
evaluated {under open boundary conditions} using 
the C-dTWA with clusters of 2 sites (orange) and the Gaussian CTWA (blue) with the same clusters. 
Here, we set $J=1/8$ and $h_x = 1$ (used as the unit of energy), the same parameters used in Ref.~\cite{wurtz2018cluster}.
For comparison, the numerically exact results obtained using iTensor~\cite{itensor} are also plotted in black. 
Note that the orange and blue curves are nearly indistinguishable at this scale.  
}
\label{fig: ising}
\end{center}
\end{figure}

Since inter-site correlations factorize in a direct product state,  
we impose the initial conditions on the cluster variables at $t=0$ as 
\begin{align}
X^{(i,j)}_{m,n}(t=0) = \zeta^{(i)}_{m} \zeta^{(j)}_{n},
\end{align}
where $\zeta^{(i)}_{m}$ describes onsite spin fluctuations.
For states $\ket{\uparrow}$ or $\ket{\downarrow}$, the $x$ and $y$ components, $\zeta^{(i)}_{x}$ and $ \zeta^{(i)}_{y}$, 
take values $+1$ or $-1$ randomly with an equal probability of $1/2$, while the $z$ components remain constant. 
Figure~\ref{fig: ising} shows the results for the quench dynamics of 
$M_z = \sum_{i=1}^{4}(-1)^{i-1}\langle {\hat \sigma}^{(i)}_{z} (t) \rangle$, 
starting from the initial state $\ket{ \psi (t=0) } = \ket{ \uparrow, \downarrow, \uparrow, \downarrow }$.
In this figure, we compare the C-dTWA results with the Gaussian CTWA results from Ref.~\cite{wurtz2018cluster} 
and find excellent agreement between these two results. 
Note that for a fair comparison, the number of trajectories is set to $20000$ for both methods.

The Gaussian Wigner function $W_{\rm Gauss} $ for the product state $\ket{ \uparrow, \downarrow, \uparrow, \downarrow }$ 
has a product form, i.e., $W_{\rm Gauss} = W_{L}W_{R}$.
Following the construction described in Ref.~\cite{wurtz2018cluster}, the distribution $W_{L(R)}$ on each cluster is given by 
\begin{align}
W_{L(R)} = &\frac{1}{{\cal Z}}\delta(x_{15} + 1)\delta(x_{3} + 1)\delta(x_{12} - 1) \nonumber \\
&\times \delta(x_1-x_{13})\delta(x_2-x_{14}) \delta(x_5-x_{10})  \nonumber \\
&\times \delta(x_4+x_{7})\delta(x_6+x_{9})\delta(x_8+x_{11}) \nonumber \\
&\times e^{-\frac{1}{2}\sum_{\alpha \in \{1,2,4,5,6,8\}}x_{\alpha}^2},
\end{align} 
where ${\cal Z}$ is the normalization factor.
The basis operators on the clusters are defined as follows: 
\begin{align}
{\hat X}_{1\;} 
&= {\hat \sigma}_{0} \otimes {\hat \sigma}_{x},\;\; {\hat X}_{2\;} = {\hat \sigma}_{0} \otimes {\hat \sigma}_{y},\;\; {\hat X}_{3\;} = {\hat \sigma}_{0} \otimes {\hat \sigma}_{z}, \nonumber \\
{\hat X}_{4\;} 
&= {\hat \sigma}_{x} \otimes {\hat \sigma}_{0},\;\; {\hat X}_{5\;} = {\hat \sigma}_{x} \otimes {\hat \sigma}_{x},\;\; {\hat X}_{6\;} = {\hat \sigma}_{x} \otimes {\hat \sigma}_{y}, \nonumber \\
{\hat X}_{7\;} 
&= {\hat \sigma}_{x} \otimes {\hat \sigma}_{z},\;\; {\hat X}_{8\;} = {\hat \sigma}_{y} \otimes {\hat \sigma}_{0},\;\; {\hat X}_{9\;} = {\hat \sigma}_{y} \otimes {\hat \sigma}_{x}, \nonumber \\
{\hat X}_{10} 
&= {\hat \sigma}_{y} \otimes {\hat \sigma}_{y},\;\; {\hat X}_{11} = {\hat \sigma}_{y} \otimes {\hat \sigma}_{z},\;\; {\hat X}_{12} = {\hat \sigma}_{z} \otimes {\hat \sigma}_{0}, \nonumber \\
{\hat X}_{13} 
&= {\hat \sigma}_{z} \otimes {\hat \sigma}_{x},\;\; {\hat X}_{14} = {\hat \sigma}_{z} \otimes {\hat \sigma}_{y},\;\; {\hat X}_{15} = {\hat \sigma}_{z} \otimes {\hat \sigma}_{z}. \nonumber
\end{align}
The distribution $W_{L(R)}$ involves six independent random numbers, and thus there are a total of twelve noise variables  
for the entire system.
In contrast, the distribution in the C-dTWA requires only eight noise variables in total.

\section{Technical details of the C-dTWA and Gaussian CTWA methods}
\label{app: manual}

In Sec.~\ref{sec: formulation}, we present a general framework for formulating the strategy of the C-dTWA 
for generic SU($N$) spin systems. In this appendix, we outline the numerical implementation of both the C-dTWA and 
Gaussian CTWA methods for the state-restricted Bose-Hubbard model. 
We particularly use the representation with $2 \times 2$ site clusters depicted in Fig.~\ref{fig: bhm2by2} and 
follow the same indexing convention for sites within the cluster.

\subsection{C-dTWA}

The first step in implementing the C-dTWA for the state-restricted bosonic model is to define an ansatz for the local phase-point 
matrices $A_{j} \in {\mathbb C}^{3 \times 3}$ at every site.
The ansatz is given by 
\begin{align}
A_{j}  
= \frac{1}{3}{\mathbb I} + \frac{1}{2}\sum_{a=1}^{8}r^{a}_{j} T^{a}.
\end{align}
In this decomposition, the base matrices in the righthand side are defined as follows: 
\begin{align}
T_1 
&= \frac{1}{\sqrt{2}}
\begin{bmatrix}
0 & 1 & 0 \\
1 & 0 & 1 \\
0 & 1 & 0
\end{bmatrix},
\;\;\;
T_2 = \frac{1}{\sqrt{2}}
\begin{bmatrix}
0 & -i & 0 \\
i & 0 & -i \\
0 & i & 0
\end{bmatrix}, \nonumber \\
T_3
&=
\begin{bmatrix}
1 & 0 & 0 \\
0 & 0 & 0 \\
0 & 0 & -1
\end{bmatrix},
\;\;\;\;\;\;
T_4
=
\begin{bmatrix}
0 & 0 & 1 \\
0 & 0 & 0 \\
1 & 0 & 0
\end{bmatrix},  \\
T_5
&=
\begin{bmatrix}
0 & 0 & -i \\
0 & 0 & 0 \\
i & 0 & 0
\end{bmatrix},
\;\;\;\;\;\;\;
T_6
= \frac{1}{\sqrt{2}}
\begin{bmatrix}
0 & -1 & 0 \\
-1 & 0 & 1 \\
0 & 1 & 0
\end{bmatrix}, \nonumber \\
T_7 
&= \frac{1}{\sqrt{2}}
\begin{bmatrix}
0 & i & 0 \\
-i & 0 & -i \\
0 & i & 0
\end{bmatrix},
\;\;\;
T_8
= \frac{1}{\sqrt{3}}
\begin{bmatrix}
-1 & 0 & 0 \\
0 & 2 & 0 \\
0 & 0 & -1
\end{bmatrix}. \nonumber
\end{align}
The weights of the Cartan generators $(T_{3}, T_{8})$, i.e., $(1,-1/\sqrt{3})$, $(0,2/\sqrt{3})$, and $(-1,-1/\sqrt{3})$, specify the low-lying 
Fock states $\ket{0}$, $\ket{1}$, and $\ket{2}$, respectively. 
The coefficients $r^{a}_{j}$ are assumed to fluctuate depending on samples, and 
distant variables are statistically uncorrelated.  
The probabilities of these fluctuations are determined via an eigenvalue decomposition of $T_{a}$, a method similar to 
quantum state tomography used in measurement-based estimation of local reduced density matrices~\cite{zhu2019generalized}.
In this approach, the probability of finding $T_{a}$ in an eigenvalue $\lambda_{n}^{(a)}$ with its eigenstate $|\phi^{(a)}_{n} \rangle$ 
is given by $p^{a}_{j}(n) = {\rm tr}\left[  \rho_{j} |\phi^{(a)}_{n} \rangle \langle \phi^{(a)}_{n} | \right]$, where $ \rho_{j}$ is the single-site 
density matrix to be reconstructed. 
For instance, in the unit-filling state $\ket{1}\bra{1}$, only four variables $r^{1}_{j}$, $r^{2}_{j}$, $r^{6}_{j}$, $r^{7}_{j}$ fluctuate, 
while the others remain constant. 
Thus, the independent noises per cluster amount to $4 l_c$, where $l_c$ is the number of sites within the cluster.
The total phase-point matrix is the direct product matrix over all $A_{j}$, expressed as $A_{\rm tot} = \otimes_j A_{j}$.
Taking the ensemble average over random realizations of the local variables successfully generates the initial density matrix, 
$\rho(t=0) = {\mathbb E}[A_{\rm tot}] = \otimes_{j}\rho_{j}$.

Mean-field trajectories in the C-dTWA are defined by the dynamical phase-point matrices ${\mathscr A}^{\rm cl}_{\mu_x,\mu_y}(t)$ 
for clusters, where $(\mu_x,\mu_y)$ denotes the real-space coordinates of the clusters. 
The global state of a trajectory, a cluster mean-field state, is specified as  
\begin{align}
{\mathscr A}_{\rm cl}(t) = \bigotimes_{\mu_x,\mu_y} {\mathscr A}^{\rm cl}_{\mu_x,\mu_y}(t).
\end{align}
The dimension of ${\mathscr A}^{\rm cl}_{\mu_x,\mu_y}(t)$ is $3^4 = 81$.
According to the general formula in Eq.~(\ref{eq: cluster_mf}),  
the evolution of the cluster state is determined by the following differential equation:   
\begin{align}
\hbar \partial_t {\mathscr A}^{\rm cl}_{\mu_x,\mu_y}(t) 
&= -i [\Gamma^{\rm intra}_{\mu_x,\mu_y}+\Phi^{\rm inter}_{\mu_x,\mu_y}, {\mathscr A}^{\rm cl}_{\mu_x,\mu_y}(t) ], \label{eq:app:eom_cdtwa}
\end{align}
where $\Gamma^{\rm intra}_{\mu_x,\mu_y}$ and $\Phi^{\rm inter}_{\mu_x,\mu_y}$ are contributions from the intra-cluster 
and inter-cluster terms, respectively. 

The intra-cluster contribution $\Gamma^{\rm intra}_{\mu_x,\mu_y}$ is derived as 
\begin{align}
\Gamma^{\rm intra}_{\mu_x,\mu_y}
= &\;\;\;\; ( H_{\rm int}^1 \otimes {\mathbb I}_{2} \otimes {\mathbb I}_{3}  \otimes {\mathbb I}_{4})^{(\mu_x,\mu_y)}\nonumber  \\
&+ ( {\mathbb I}_{1} \otimes H_{\rm int}^2 \otimes {\mathbb I}_{3}  \otimes {\mathbb I}_{4})^{(\mu_x,\mu_y)} \nonumber  \\
&+ ( {\mathbb I}_{1} \otimes  {\mathbb I}_{2}  \otimes H_{\rm int}^3  \otimes {\mathbb I}_{4})^{(\mu_x,\mu_y)} \nonumber  \\
&+ ( {\mathbb I}_{1} \otimes  {\mathbb I}_{2}  \otimes {\mathbb I}_{3} \otimes H_{\rm int}^4)^{(\mu_x,\mu_y)}  \nonumber  \\
&- J ( a^{\dagger}_{1} \otimes a_{2}  \otimes {\mathbb I}_{3} \otimes {\mathbb I}_{4})^{(\mu_x,\mu_y)} \nonumber  \\
&- J ( a_{1} \otimes a^{\dagger}_{2}  \otimes {\mathbb I}_{3} \otimes {\mathbb I}_{4})^{(\mu_x,\mu_y)} \nonumber  \\
&- J (  {\mathbb I}_{1}  \otimes a_{2}  \otimes {\mathbb I}_{3} \otimes a^{\dagger}_{4})^{(\mu_x,\mu_y)} \nonumber  \\
&- J (  {\mathbb I}_{1}  \otimes a^{\dagger}_{2}  \otimes {\mathbb I}_{3} \otimes a_{4})^{(\mu_x,\mu_y)}  \nonumber  \\
&- J (  {\mathbb I}_{1}  \otimes {\mathbb I}_{2} \otimes a^{\dagger}_{3}  \otimes a_{4})^{(\mu_x,\mu_y)} \nonumber  \\
&- J (  {\mathbb I}_{1}  \otimes {\mathbb I}_{2} \otimes a_{3}  \otimes a^{\dagger}_{4})^{(\mu_x,\mu_y)}  \nonumber  \\
&- J ( a_{1} \otimes {\mathbb I}_{2}  \otimes a^{\dagger}_{3}  \otimes {\mathbb I}_{4})^{(\mu_x,\mu_y)}  \nonumber  \\
&- J ( a^{\dagger}_{1} \otimes {\mathbb I}_{2}  \otimes a_{3}  \otimes {\mathbb I}_{4})^{(\mu_x,\mu_y)}, \nonumber 
\end{align}
where $H_{\rm int}^{i_{\rm c}}$ and $a_{i_{\rm c}}$ represent the two-body interaction term in Eq.~(\ref{eq:H_bose}) 
and the annihilation operator of bosons, respectively, at the $i_{\rm c}$th site within the cluster located at $(\mu_x,\mu_y)$.
The inter-cluster term $\Phi^{\rm inter}_{\mu_x,\mu_y}$ is given by  
\begin{align}
\Phi^{\rm inter}_{\mu_x,\mu_y}
&= -J (a_{1} \otimes {\mathbb I}_{2}  \otimes {\mathbb I}_{3}  \otimes {\mathbb I}_{4})^{(\mu_x,\mu_y)}  \nonumber \\
&\;\;\;\;\;\;\;\;\;\;\;\; \times \langle ({\mathbb I}_{2}  \otimes {\mathbb I}_{2}  \otimes {a}^{\dagger}_{3}  \otimes {\mathbb I}_{4})^{(\mu_x-1,\mu_y)}  \rangle \nonumber \\
&- J (a_{1} \otimes {\mathbb I}_{2}  \otimes {\mathbb I}_{3}  \otimes {\mathbb I}_{4})^{(\mu_x,\mu_y)}  \nonumber \\
&\;\;\;\;\;\;\;\;\;\;\;\; \times  \langle ({\mathbb I}_{1}  \otimes {a}^{\dagger}_{2}  \otimes {\mathbb I}_{3}  \otimes {\mathbb I}_{4})^{(\mu_x,\mu_y-1)}  \rangle \nonumber \\
&- J ( {\mathbb I}_{1} \otimes a_{2} \otimes {\mathbb I}_{3}  \otimes {\mathbb I}_{4})^{(\mu_x,\mu_y)}  \nonumber \\
&\;\;\;\;\;\;\;\;\;\;\;\; \times  \langle ({a}^{\dagger}_{1} \otimes  {\mathbb I}_{2}  \otimes {\mathbb I}_{3}  \otimes {\mathbb I}_{4})^{(\mu_x,\mu_y+1)}  \rangle \nonumber \\
&- J ( {\mathbb I}_{1} \otimes a_{2} \otimes {\mathbb I}_{3}  \otimes {\mathbb I}_{4})^{(\mu_x,\mu_y)}  \nonumber \\
&\;\;\;\;\;\;\;\;\;\;\;\; \times  \langle ({\mathbb I}_{1} \otimes  {\mathbb I}_{2}  \otimes {\mathbb I}_{3}  \otimes {a}^{\dagger}_{4} )^{(\mu_x-1,\mu_y)}  \rangle \nonumber \\
&-  J ( {\mathbb I}_{1} \otimes {\mathbb I}_{2}  \otimes a_{3} \otimes {\mathbb I}_{4})^{(\mu_x,\mu_y)}  \nonumber \\
&\;\;\;\;\;\;\;\;\;\;\;\; \times  \langle ({a}^{\dagger}_{1} \otimes  {\mathbb I}_{2}  \otimes {\mathbb I}_{3}  \otimes {\mathbb I}_{4} )^{(\mu_x+1,\mu_y)}  \rangle \nonumber \\
&- J ( {\mathbb I}_{1} \otimes {\mathbb I}_{2}  \otimes a_{3} \otimes {\mathbb I}_{4})^{(\mu_x,\mu_y)}  \nonumber \\
&\;\;\;\;\;\;\;\;\;\;\;\; \times  \langle ({\mathbb I}_{1} \otimes  {\mathbb I}_{2}  \otimes {\mathbb I}_{3}  \otimes {a}^{\dagger}_{4} )^{(\mu_x,\mu_y-1)}  \rangle \nonumber \\
&-  J ( {\mathbb I}_{1} \otimes {\mathbb I}_{2}  \otimes {\mathbb I}_{3} \otimes  a_{4} )^{(\mu_x,\mu_y)}  \nonumber \\
&\;\;\;\;\;\;\;\;\;\;\;\; \times  \langle ({\mathbb I}_{1} \otimes  {\mathbb I}_{2}  \otimes  {a}^{\dagger}_{3} \otimes {\mathbb I}_{4} )^{(\mu_x,\mu_y+1)}  \rangle \nonumber \\
&- J ( {\mathbb I}_{1} \otimes {\mathbb I}_{2}  \otimes  {\mathbb I}_{3} \otimes  a_{4} )^{(\mu_x,\mu_y)}  \nonumber \\
&\;\;\;\;\;\;\;\;\;\;\;\; \times  \langle ({\mathbb I}_{1} \otimes  {a}^{\dagger}_{2}  \otimes  {\mathbb I}_{3}  \otimes {\mathbb I}_{4} )^{(\mu_x+1,\mu_y)}  \rangle \nonumber \\
&+  {\rm H.c.},  \nonumber
\end{align}
where the expectation values $\langle \cdots \rangle$ are evaluated with respect to the state at time $t$ of the neighboring clusters 
connected to the cluster at $(\mu_x,\mu_y)$.
The bosonic operators are now represented as the three-dimensional matrices: 
\begin{align}
a_{i_{\rm c}} = 
\begin{pmatrix}
0 & 1 & 0 \\
0 & 0 & \sqrt{2} \\
0 & 0 & 0
\end{pmatrix},
a^{\dagger}_{i_{\rm c}} = 
\begin{pmatrix}
0 & 0 & 0 \\
1 & 0 & 0 \\
0 & \sqrt{2}  & 0
\end{pmatrix},
H_{\rm int}^{i_{\rm c}} = 
\begin{pmatrix}
0 & 0 & 0 \\
0 & 0 & 0\\
0 & 0 & U
\end{pmatrix}. \nonumber 
\end{align}
The differential equation for the $81 \times 81$ matrices ${\mathscr A}^{\rm cl}_{\mu_x,\mu_y}(t)$ is solved using 
the explicit fourth-order Runge-Kutta algorithm. 
Initially, the cluster matrices $ {\mathscr A}^{\rm cl}_{\mu_x,\mu_y}(t) $ are factorized, i.e., 
${\mathscr A}^{\rm cl}_{\mu_x,\mu_y}(t = 0)  = \bigotimes_{j \in S_{\mu_x,\mu_y}}A_{j} $, where $A_{j}$ are randomly sampled 
matrices and $S_{\mu_x,\mu_y}$ denotes the set of sites within the cluster. 
These initial conditions are used to solve the differential equation in Eq.~(\ref{eq:app:eom_cdtwa}).

It is worth noting that the formalism introduced here can be straightforwardly applied to open quantum systems described by the 
Lindblad master equation~\cite{nagao2023semiclassical}.
A key feature of this approach is the local linearization of dissipative terms expressed by Kraus operators, which cannot 
be achieved with the Schwinger-boson-based approach underlying the Gaussian CTWA.

\subsection{Gaussian CTWA}

The first step in performing the Gaussian CTWA, originally introduced in Ref.~\cite{wurtz2018cluster}, is to construct cluster spin 
operators directly from individual spin operators. 
For the three-state restricted Bose-Hubbard model involving four-site clusters, we define a list of all tensor products of SU(3) 
spin operators as 
${\hat X}_{\alpha}^{(\mu_x,\mu_y)} \equiv {\hat s}^{a}_{1} \otimes {\hat s}^{b}_{2} \otimes {\hat s}^{c}_{3} \otimes {\hat s}^{d}_{4}$
for $\alpha \in \{1,2,\cdots, D\}$, 
where the trivial unit operator is omitted in this definition. 
The individual spin operators satisfy the commutation relations 
$[{\hat s}^{a}_{j} ,{\hat s}^{b}_{j'}] = i \sum_{c=1}^{8}\epsilon_{abc} {\hat s}^{c}_{j} \delta_{j,j'}$.
To generate semiclassical time evolution, we approximate the cluster operators ${\hat X}_{\alpha}^{(\mu_x,\mu_y)} $ using 
the corresponding differential Bopp operators in phase space~\cite{polkovnikov2010phase}:  
\begin{align}
{\hat X}_{\alpha}^{(\mu_x,\mu_y)} \rightarrow x_{\alpha}^{(\mu_x,\mu_y)} + \frac{i}{2} \sum_{\beta,\gamma=1}^{D} f_{\alpha\beta\gamma} x_{\gamma}^{(\mu_x,\mu_y)} \frac{\partial}{\partial x_{\beta}^{(\mu_x,\mu_y)}}, \label{eq:app:bopp}
\end{align}
where $f_{\alpha\beta\gamma}$ are the structure constants of an SU(81) Lie group, which can be straightforwardly calculated 
from the commutation relations of ${\hat X}_{\alpha}^{(\mu_x,\mu_y)}$.
The mapping in Eq.~(\ref{eq:app:bopp}) is a crucial step in the standard CTWA and can be systematically derived 
by performing a leading-order truncation of bosonic bilinear representation of SU($N$) generators~\cite{wurtz2018cluster}.

Within the validity of Eq.~(\ref{eq:app:bopp}), the dynamics of canonical cluster variables is generated by the Hamilton equation: 
\begin{align}
{\dot x}^{(\mu_x,\mu_y)}_{\alpha} = -i\{x^{(\mu_x,\mu_y)}_{\alpha},H_{W}\}_{\rm P.B.}, \label{eq:app:hamilton_cluster}
\end{align}
where $H_{W}$ is the Weyl symbol of the Hamiltonian, and 
$$
\{A,B\}_{\rm P.B.} = \sum_{\mu_x,\mu_y}\sum_{\alpha,\beta,\gamma} i f_{\alpha \beta \gamma} x^{(\mu_x,\mu_y)}_{\gamma} \frac{\partial A}{\partial x^{(\mu_x,\mu_y)}_{\alpha}}  \frac{\partial B}{\partial x^{(\mu_x,\mu_y)}_{\beta}}
$$ 
is the Poisson bracket. 
Note that this equation is a projection of Eq.~(\ref{eq:app:eom_cdtwa}) onto the specific basis of ${\hat X}_{\alpha}^{(\mu_x,\mu_y)}$. 
In the case of isolated unitary dynamics governed by the von-Neumann equation, there is essentially no difference between 
applying Eq.~(\ref{eq:app:eom_cdtwa}) or Eq.~(\ref{eq:app:hamilton_cluster}).
However, when there exists a finite coupling between the system and its environment, higher order derivative terms, which 
are truncated in Eq.~(\ref{eq:app:bopp}), contribute non-negligible effects to the dynamics of the system, whereas they effects 
are difficult to accurately integrate in the Schwinger-boson approach~\cite{nagao2023semiclassical}.

Next, we introduce the sampling probability using a Gaussian ansatz. 
For simplicity, we omit the spatial indices on clusters in the following. 
We define a Gaussian Wigner function for fluctuating cluster variables as
\begin{align}
W(\{x\}) = \frac{1}{\cal Z} \exp\left[ -\frac{1}{2}(x_\alpha - \rho_\alpha) \Sigma^{-1}_{\alpha \beta} (x_\beta - \rho_\beta) \right], \label{eq: gauss_wigner}
\end{align}
where ${\cal Z}$ is the normalization factor~\cite{wurtz2018cluster}. 
This distribution is uniquely specified by two free parameters, $\rho_{\alpha}$ and $\Sigma_{\alpha\beta}$, which are real-valued.
These parameters are determined by conditions on the first and second moments, i.e., 
\begin{align}
{\rm Tr}[{\hat \rho}{\hat X}_{\beta} ] 
&= \int \left[ \prod_{\alpha=1}^{D}d x_{\alpha}  \right] x_{\beta} W(\{x\}) = \mu_{\beta}, \\
{\rm Tr}[{\hat \rho} \{ {\hat X}_{\beta}, {\hat X}_{\gamma} \} ] 
&= 2 \int \left[ \prod_{\alpha=1}^{D}d x_{\alpha}  \right] x_{\beta} x_{\gamma} W(\{x\}).
\end{align}
The $D\times D$ matrix $({\bf \Sigma})_{\alpha,\beta} = \Sigma_{\alpha \beta}$ with ${\alpha,\beta \in \{1,\cdots,D\}}$ 
in Eq.~(\ref{eq: gauss_wigner}) is the covariance matrix of the Gaussian distribution, defined by 
$\Sigma_{\alpha \beta} = \frac{1}{2}{\rm Tr}[{\hat \rho}\{ {\hat X}_{\alpha}, {\hat X}_{\beta} \} ] - \mu_{\alpha} \mu_{\beta} $, 
where $\{A,B\}=AB+BA$.
For the $2\times2$ site clusters used in our study, the upper bound $D$ is set to be $D=3^4 - 1 = 6560$. 
Since ${\bf \Sigma}$ is real and symmetric, it can be diagonalized with an orthogonal matrix $R$.
The eigenvalues $\sigma^2_\alpha$ for ${\alpha\in {1,\cdots, D}}$ represent the variances of decoupled normal modes, 
as given by the spectral decomposition: 
\begin{align}
{\Sigma}_{\alpha \beta}  = \sum_{\gamma = 1}^{D}R_{\alpha \gamma} (\sigma^2_{\gamma} ) R^{t}_{\gamma \beta}\; \Rightarrow \; {\Sigma}^{-1}_{\alpha \beta}  = \sum_{\gamma = 1}^{D} R_{\alpha \gamma} \frac{1}{\sigma^2_{\gamma}} R^{t}_{\gamma \beta}.
\end{align}
Random numbers $x_{\alpha}$ from the Gaussian Wigner distribution are generated through the linear transformation 
$x_{\alpha} = \sum_{\beta}R_{\alpha \beta}\sigma_{\beta}y_{\beta} + \rho_{\alpha}$, where the random numbers $y_{\beta}$ 
are drawn from a normal distribution with unit variance. 
We have numerically identified that only 160 eigenvalues are nonzero for the $\ket{1111}$ state, 
corresponding to the unit-filling Mott insulating state for four sites, with negligible errors on the order of 
${\cal O}(10^{-8})$, as accumulated after running a LAPACK routine. 
The number of independent noises in this Gaussian approach is substantially larger than in the C-dTWA, 
which requires only sixteen per cluster.

\bibliography{ref}

\end{document}